# Effect of non-local electron conductivity on power absorption and plasma density profiles in low pressure inductively coupled discharges.

*Badri Ramamurthi and Demetre J. Economou*[1]
*Plasma Processing Laboratory, Department of Chemical Engineering*
*University of Houston, Houston, TX 77204-4004*

Igor D. Kaganovich[2]
*Plasma Physics Laboratory, Princeton University, Princeton, NJ 08543*

A self-consistent 1-D model was developed to study the effects of non-local electron conductivity on power absorption and plasma density profiles in a planar inductively coupled argon discharge at low pressures (≤ 10 mTorr). The model consisted of three modules: (1) an electron energy distribution function (EEDF) module to compute the non-Maxwellian EEDF, (2) a non-local electron kinetics module to predict the non-local electron conductivity, RF current, electric field and power deposition profiles in the non-uniform plasma, and (3) a heavy species transport module to solve for the ion density and velocity profiles as well as the metastable density. Results using the non-local electron conductivity model were compared with predictions of a local theory (Ohm's law), under otherwise identical conditions. The RF current, electric field, and power deposition profiles were very different, especially at 1 mTorr for which the effective electron mean free path was larger than the skin depth. However, the plasma density profiles were almost identical (within 10%) for the same *total* power deposition in the plasma. This result suggests that, for computing plasma density profiles, a local conductivity model (Ohm's law), with much reduced computational expense, may be employed even in the non-local regime.

## 1.0 INTRODUCTION

Inductively coupled plasma (ICP) sources can produce a high-density, uniform plasma in a low pressure gas without the need for external magnetic fields [1,2,3,4,5]. Such sources are used extensively for etching and deposition of thin films in microelectronics manufacturing.

At higher pressures (above ≈ 20 mTorr), electrons in an ICP discharge are heated by collisional dissipation of wave energy. However, both experimental and theoretical results in lower pressure discharges, indicate that power deposition involves a collisionless electron heating mechanism [6,7]. It has been suggested that in both planar [8] and solenoidal [4] ICP discharges, the collisionless heating mechanism is a "warm plasma" effect analogous to the anomalous skin effect in metals.

The anomalous skin effect in gas discharges was first studied analytically by Weibel [9] for a semi-infinite plasma with uniform electron density. Further analytical work was performed for a non-uniform density, semi-infinite plasma with a "diffuse boundary" by Liberman et al. [10], and for an infinite plasma with a "diffuse boundary" by Dikman et al. [11]. Early experimental investigation of the skin effect was performed by Demirkhanov et al. [12] for a cylindrical "ring" discharge.

The anomalous skin effect in 1-D (slab geometry) *bounded* plasmas has been studied theoretically and experimentally in [13,14] for a symmetric power source (a current sheet on either side of the slab), and in [15,16] for an asymmetric source. An interesting effect associated with bounded plasmas is the possible resonance between the wave frequency and the motion of electrons bouncing between the walls. This can lead to enhanced heating [15,17,18]. However, as shown in Ref. 19, if the electron elastic collision frequency is too small, heating can actually decrease. Most theoretical results reported thus far for a bounded plasma assume a uniform density plasma, where the electrostatic potential well is flat in the plasma and infinite at the wall (to simulate the existence of sheaths). In this square potential well, electrons are reflected back into the plasma only at the discharge walls. In a realistic non-uniform plasma, however, the electron turning points will depend on the electron total (kinetic plus potential) energy and the actual shape of the potential well, i.e., low total energy electrons bounce back at locations within the plasma. Although theoretical treatments of non-uniform slab plasmas have been reported [10], detailed self-consistent simulations related to such plasmas are lacking. A review of classical and recent works on the anomalous skin effect in plasmas was made in [20].

The present article presents a self-consistent simulation of non-local electron kinetics and heavy species transport in a 1-D slab (bounded) non-uniform plasma with a symmetric power source. An argon discharge is studied incorporating both electron impact reactions and metastable chemistry. Non-local effects on both power deposition and plasma density *profiles* are of particular interest. It is shown that non-local behavior strongly influences the RF field and power deposition profiles while having a negligible impact on plasma density profiles for the same *total* power.

## 2.0 MODEL DESCRIPTION

A schematic of a 1-D parallel plate symmetric discharge (plate separation L) is shown in Fig. 1. Current sheets (not shown) on either side of the plasma, driven by a radio frequency (RF) source, generate a transverse RF field $E_y$ heating the plasma electrons. The RF field amplitude at the plasma edges is $E_0$; this value is set by the magnitude of the RF current and directly affects the total power deposited in the plasma. The RF field is attenuated by power transfer to the plasma electrons. Most of the power is deposited near the edge in what is called the "skin layer." Predicting the RF field and power deposition profiles with account of the thermal motion of electrons in a self-consistent manner is an important part of this work.

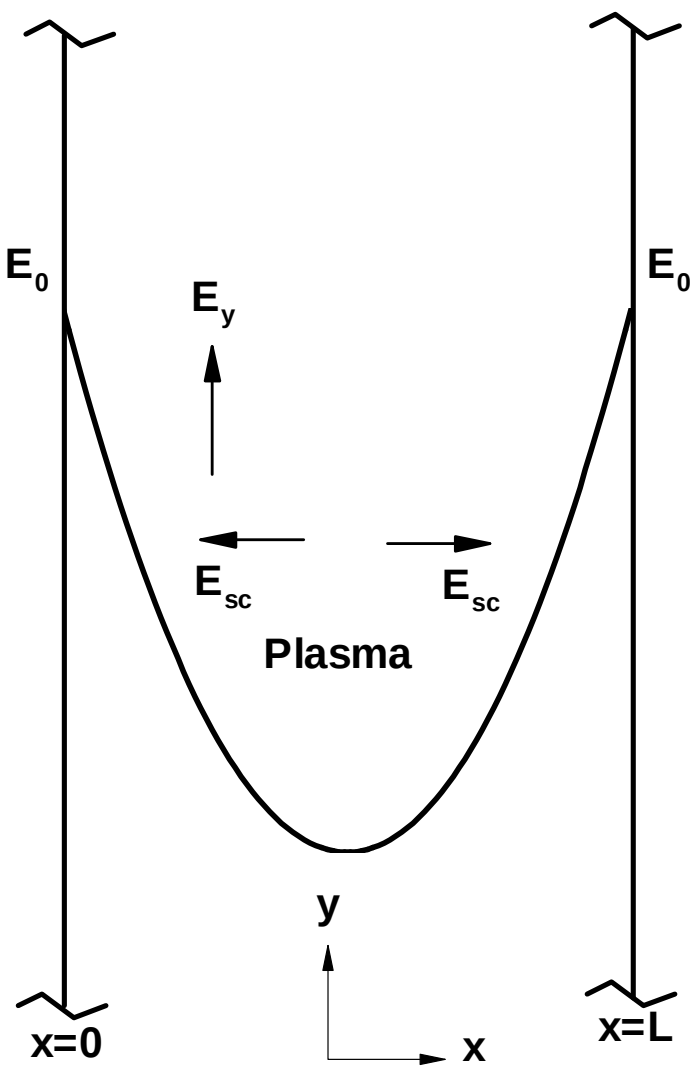

**Figure 1**: Schematic of a one-dimensional plasma slab of length $L$ powered by a symmetric inductively coupled source. The RF current source (not shown) results in an RF field in the transverse direction, $E_y$. The value of the field at the edges, $E_0$, is determined by the desired power deposition in the plasma. A space charge field $E_{sc}$ develops in the $x$-direction to confine electrons.

An electrostatic (space charge) field $E_{sc}(x)$ in the $x$-direction develops to confine electrons in the plasma and equalize the electron and ion current to the walls. The electron potential energy $\varphi(x)$ corresponding to this field is shown schematically in Fig. 2. Electrons with sufficiently low total (x-kinetic plus potential) energy will be reflected by this potential well. The reflection points $x_1^*$ and $x_2^*$ for an electron with total energy ε are shown in Fig. 2. Thus, low energy electrons are confined near the discharge center, but higher energy electrons can reach further towards the walls. The sheath near the physical boundaries was not accounted for explicitly. Because of the high plasma density the sheath is only 100s of μm thick. Thus, the location of the sheath edge is essentially at the physical boundary, and the plasma approximation $n_i=n_e$ was applied to the whole domain. An infinite potential barrier was assumed for the sheath. Electrons with total energy higher than the potential at the sheath edge $\varphi_{sh}$ were assumed to reflect at the physical boundary, the underlying assumption being that the electron current to the wall was considered to be negligible.

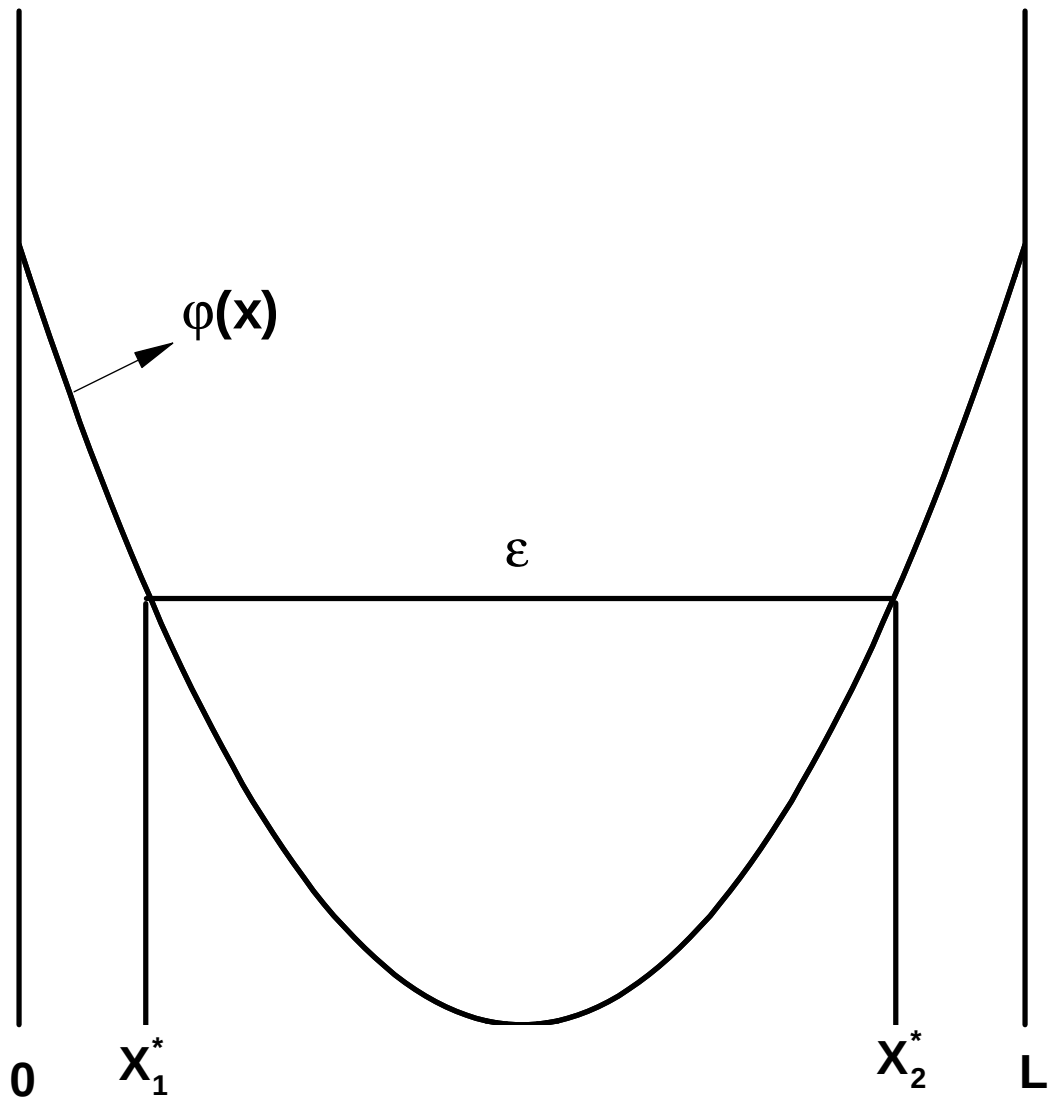

**Figure 2**: Schematic of the electron potential energy profile $\varphi(x)$ due to the electrostatic field in the plasma slab. An electron with total ($x$-kinetic plus potential) energy $\varepsilon$ will reflect back at points $x_1$* and $x_2$* (turning points).

Since non-local behavior is a warm plasma effect, kinetic treatment of electron transport is necessary. When electrons are warm enough to be transported out of the "skin layer" during an RF cycle, power is said to be deposited non-locally. In a sense, the current at a given location is influenced by the field at all other locations. In contrast, in the local case, the current at a given location only depends on the field at that particular point (Ohm's law). Non-locality is typically characterized by the parameter $l / \delta_0$, where $l = V_T / \sqrt{\omega^2 + \nu^2}$ is an "effective" electron mean free path, and $\delta_0$ is derived from the classical skin depth,

$$\delta_0 = \frac{c}{\omega_p}\left(1 + \frac{\nu^2}{\omega^2}\right)^{1/4}. \qquad (1)$$

Here $V_T = \sqrt{2eT_e / m}$ is the most probable electron speed, $\omega_p$ is the electron plasma frequency ($\omega_p = \sqrt{e^2 n_e / m\varepsilon_0}$), $\omega$ is the RF frequency, $c$ is the speed of light in vacuum, $\upsilon$ is the electron momentum-transfer collision frequency, $T_e$ is the electron temperature (in V), for a Maxwellian distribution function, and $m$ is the electron mass. The non-locality parameter mentioned above is strictly applicable to a semi-infinite plasma. Nevertheless, it is used here even for a finite plasma slab to aid discussion of results.

For pressures typically smaller than 10 mTorr for argon, the electron energy distribution function (EEDF) can be non-Maxwellian [21]. Hence, for accurate calculation of discharge characteristics at low pressures, the EEDF needs to be computed self-consistently. The following section describes a model for computing the EEDF for a low pressure argon plasma in which collisionless electron heating can be dominant [22, 23]. Only a sketchy outline of the model is given here. Details are left for another paper [23].

### 2.1 Electron Energy Distribution Function (EEDF) Module

The Boltzmann equation for the EEDF $f$ (assuming a spatial dependence only in the x-direction) can be written as

$$\frac{\partial f}{\partial t} + v_x \frac{\partial f}{\partial x} - \frac{eE_{sc}(x)}{m}\frac{\partial f}{\partial v_x} - \frac{eE_y e^{i\omega t}}{m}\frac{\partial f}{\partial v_y} = S(f), \quad (2)$$

where $E_{sc}(x)$ is the electrostatic field, and $S(f)$ represents the sum of electron-atom (elastic and inelastic) and electron-electron collisions. For small deviations from a stationary EEDF $f_0$, one can write $f = f_0(x, v_x, v_y, v_z) + f_1(x, v_x, v_y, v_z, t)$, assuming that the relaxation time of $f_0$ is small compared to the RF period. Substituting for $f$ in Eq. (2), assuming a harmonic dependence of $f_1$, and integrating over the RF period, one finally obtains an equation for $f_0$ in terms of the total energy $\varepsilon_x = mv_x^2 / 2e + \varphi(x)$.

$$\frac{\partial}{\partial \varepsilon}\left( D_\varepsilon(\varepsilon) \frac{\partial f_0}{\partial \varepsilon} \right) = \overline{S(f_0)} \quad (3)$$

The electron potential energy $\varphi(x) > 0$ everywhere, except that $\varphi(x) = 0$ at the discharge center. The energy diffusion coefficient $D_\varepsilon(\varepsilon)$ is given by [23]

$$D_\varepsilon(\varepsilon) = \frac{\pi}{8}\left(\frac{2e}{m}\right)^{3/2} \sum_{n=-\infty}^{\infty} \int_0^{\varepsilon} d\varepsilon_x \frac{\nu(\varepsilon - \varepsilon_x)\left|E_{yn}(\varepsilon_x)\right|^2}{\Omega_b(\varepsilon_x)\left(\left[\Omega_b(\varepsilon_x) n - \omega\right]^2 + \nu^2\right)} \quad (4)$$

where $\Omega_b$ is the frequency of an electron bouncing in the potential well (bounce frequency). For $\nu >> \Omega_b$, the energy diffusion coefficient reduces to the collisional limit [24].

The right hand side of Eq. (3) denotes the space- and bounce time-average of electron-atom and electron-electron collisions [24]. Collisions considered in the model were elastic electron-atom collisions, inelastic electron-atom collisions (ground-state ionization, excitation, and metastable ionization), and electron-electron collisions.

Boundary conditions for Eq. (3) were specified for large energies assuming that both $f_0$ and $\partial f_0 / \partial \varepsilon$ are "small" (note that both cannot be zero as the integration of the discretized form of Eq. (3) would not proceed). The exact values for $f_0$ and $\partial f_0 / \partial \varepsilon$ do not matter as the distribution function was finally normalized such that

$$\int_0^{\infty} \sqrt{\varepsilon} f_0(\varepsilon) d\varepsilon = 1 \quad (5)$$

Eq. (3) was solved as an initial value problem, starting with values for $f_0$ and $\partial f_0 / \partial \varepsilon$ at $\varepsilon = 75$ V and marching backwards to $\varepsilon = 0.01$ V, using a fourth order Runge-Kutta scheme.

### 2.2 Non-local Electron Kinetics (NLEK) Module

Maxwell's equations can be reduced to a single scalar equation for the transverse electric field $E_y$,

$$\frac{d^2 E_y}{dx^2} = i\omega\mu_0 J_y \quad (6)$$

where the displacement current has been neglected since the field frequency $\omega \ll \omega_p$, the electron plasma frequency.

In the local limit when $l \ll \delta_0$, Ohm's law is valid and the current $J_y$ can be written as

$$J_y(x) = \sigma(x) E_y(x), \quad (7)$$

where the local conductivity $\sigma(x)$ for a non-Maxwellian EEDF $f_0$ is defined by [24]

$$\sigma(x) = \frac{-2e^2}{3m} \int_{\varphi(x)}^{\infty} \frac{[\varepsilon - \varphi(x)]^{3/2}}{\nu + i\omega} \frac{\partial f_0(\varepsilon)}{\partial \varepsilon} d\varepsilon. \quad (8)$$

In the non-local limit of $l \gg \delta_0$, Ohm's law for current density is not valid any longer and needs to be replaced by a non-local conductivity operator. The relationship between current density and electric field in the non-local limit for a non-Maxwellian EEDF can be written as [10,11]

$$J_y(x) = \frac{e^2 n_{e0}}{2m}\sqrt{\frac{m}{2e}} \times \left( \int_0^x G(x, x') E_y(x') dx' + \int_x^L G(x', x) E_y(x') dx' \right) \quad (9)$$

Note that the current at a particular location $x$ depends on the value of the electric field $E_y$ throughout the plasma. The current also depends on the potential profile coming in through the conductivity kernel $G(x,x')$ (see Eq. 12 below). Substituting for $J_y$ in Eq. (6), one obtains

$$\frac{d^2 E_y}{dx^2} = i\left(\frac{\omega_{p0}}{c}\right)^2 \frac{\omega}{2}\sqrt{\frac{m}{2e}} \left( \int_0^x G(x, x') E_y(x') dx' + \int_x^L G(x', x) E_y(x') dx' \right) \quad (10)$$

where $\omega_{p0} = (e^2 n_{e0} / m\, \varepsilon_0)^{1/2}$ is the electron plasma frequency evaluated using the peak electron density $n_{e0}$ (at the discharge center). The boundary conditions for $E_y$ were $E_y(0) = E_y(L) = E_0$. The time-average power deposition profile was then computed as

$$P(x) = \frac{1}{2}\mathrm{Re}\left(J_y(x)E_y^*(x)\right) \tag{11}$$

where $E_y^*(x)$ is the complex conjugate of $E_y(x)$, and Re is the real part of the quantity in parenthesis.

The functional form of *G(x,x')* was modified from that given in [11] to account for a non-Maxwellian EEDF [23]

$$G(x',x) = 2\int_0^\infty A(x',x,v)\frac{\Gamma(\varepsilon)}{\left(v^2 + 2e|\varphi(x) - \varphi(x')|/m\right)^{1/2}}dv$$

$$AA(x',x,v) = \frac{\cosh(\Phi(x_1^*,x))\cosh(\Phi(x',x_2^*))}{\sinh(\Phi(x_1^*,x_2^*))} \tag{12}$$

where

$$\Gamma(\varepsilon) = \int_\varepsilon^\infty f_0(\varepsilon')d\varepsilon' \tag{13}$$

and $\varepsilon = mv^2/2e + [\varphi(x) + \varphi(x') + |\varphi(x) - \varphi(x')|]/2$. In Eq. (12), the phase change of an electron with respect to the RF field as it moves from point $x_0$ to $x$ is given by

$$\Phi(x_0,x) = \int_{x_0}^{x}\frac{i\omega + \nu}{\sqrt{2e(\varepsilon - \varphi(x'))/m}}dx'. \tag{14}$$

Once the isotropic EEDF $f_0$ can be obtained by solving Eq. (3), Eq. (10) along with Eqs. (12)-(14) can be used to compute the RF field profile in the discharge.

### 2.3 Heavy Species Transport (HST) Module

The heavy species transport module solves for the ion density and velocity profiles as well as the metastable species density. Since the extremely thin sheaths were not included in the simulation, the quasi-neutrality constraint was imposed, and the location of the plasma-sheath boundary was taken to be essentially at the wall. The power deposition in the plasma is implicitly coupled to the ion equation through ionization which depends on impact cross-sections of electron-atom collisions and the EEDF. Since the drift-diffusion approximation for ions is questionable at pressures below about 10 mTorr, a momentum equation was solved to compute $Ar^+$ velocity. The metastable Ar* species density is quite uniform at pressure below about 10 mTorr. Hence a spatially average (0-D) model was used to determine Ar*. Due to symmetry, only half the domain (*0≤x≤L/2*) was considered.

The continuity equation for $Ar^+$ ions can be written as

$$\frac{\partial n_+}{\partial t} + \frac{\partial(n_+u_+)}{\partial x} = R_i + R_{mi} + R_{mp} \tag{15}$$

where the reactions on the right hand side represent ground-state ionization, metastable (step-wise) ionization and metastable pooling respectively (Table I). Linear extrapolation was employed to compute the density at the boundary and the density gradient at the center of the discharge was set to zero.

**Table I**: Reactions used in the argon discharge simulation [25]. $\Delta H_j$ is electron energy loss (positive value) or gain (negative value) upon collision.

| No. | Process | Symbol | Reaction | $\Delta H_j$ (eV) |
|---|---|---|---|---|
| R1 | Ground state excitation | $R_{ex}$ | $Ar + e \rightarrow Ar^* +$ | 11.6 |
| R2 | Ground state ionization | $R_i$ | $Ar + e \rightarrow Ar^+ +$ | 15.8 |
| R3 | Step-wise ionization | $R_{mi}$ | $Ar^* + e \rightarrow Ar^+$ | 4.2 |
| R4 | Superelastic collisions | $R_{sc}$ | $Ar^* + e \rightarrow Ar +$ | -11.6 |
| R5 | Metastable quenching | $R_{mq}$ | $Ar^* + e \rightarrow Ar^r$ | |
| R6 | Metastable pooling | $R_{mp}$ | $Ar^* + Ar^* \rightarrow Ar$ | |
| R7 | Two-body quenching | $R_{2q}$ | $Ar^* + Ar \rightarrow 2Ar$ | |
| R8 | Three-body quenching | $R_{3q}$ | $Ar^* + 2Ar \rightarrow Ar$ | |

To determine ion velocity, one starts with the momentum equation,

$$\frac{\partial(n_+u_+)}{\partial t} + \frac{\partial(n_+u_+u_+)}{\partial x} = \frac{en_+}{m_+}\frac{\partial\varphi}{\partial x} - \frac{eT_+}{m_+}\frac{\partial n_+}{\partial x} - \nu_+(u_+)n_+u_+ \tag{16}$$

Using the ion continuity equation, Eq. (16) can be simplified to a new equation for ion velocity,

$$\frac{\partial u_+}{\partial t} + u_+ \frac{\partial u_+}{\partial x} = \frac{e}{m_i}\frac{\partial \varphi}{\partial x} - \frac{eT_i}{m_i}\frac{\partial(\ln n_+)}{\partial x} - \nu_i(u_+)u_+ - \frac{(R_i + R_{mi} + R_{mp})}{n_+}u_+, \quad (17)$$

where $n_+$ and $u_+$ are positive ion density and velocity respectively, and $m_+$ is the ion mass. Potential $\varphi(x)$ was obtained from the EEDF module as follows. The electron density can be computed from $f_0$ as

$$n_e(x) = n_+(x) = \int_{\varphi(x)}^{\infty} \sqrt{\varepsilon - \varphi(x)} f_0(\varepsilon) d\varepsilon \quad (18)$$

where $(\varepsilon - \varphi(x))$ is the electron kinetic energy.

Differentiating Eq. (18) with respect to $x$ on both sides yields

$$\frac{\partial \varphi}{\partial x} = -T_{eff}(x)\frac{\partial(\ln n_+)}{\partial x} \quad (19)$$

where $T_{eff}(x)$, which is also known as the "screening temperature" [26] was defined by

$$T_{eff}(x) = \frac{2\int_{\varphi(x)}^{\infty} \sqrt{\varepsilon - \varphi(x)} f_0(\varepsilon) d\varepsilon}{\int_{\varphi(x)}^{\infty} \frac{f_0(\varepsilon) d\varepsilon}{\sqrt{\varepsilon - \varphi(x)}}}. \quad (20)$$

Potential $\varphi(x)$ was then computed by integrating Eq. (19)

$$\varphi(x) = \int_{x}^{L/2} T_{eff}(x)\frac{\partial(\ln n_+ / n_{+0})}{\partial x} dx, \quad (21)$$

where the potential value at the discharge center $\varphi(L/2)$ was chosen to be zero. Here, $n_{e0}$ is the electron density at the center (to be determined self-consistently).

Substituting Eq. (19) in Eq. (17), the final form of the equation for ion velocity is obtained,

$$\frac{\partial u_+}{\partial t} + u_+ \frac{\partial u_+}{\partial x} = -\frac{eT_{eff}(x)}{m_+}\frac{\partial(\ln n_+)}{\partial x} - \frac{eT_+}{m_+}\frac{\partial(\ln n_+)}{\partial x} - \nu_+(u_+)u_+ - \frac{(R_i + R_{mi} + R_{mp})}{n_+}u_+, \quad (22)$$

The second term on the RHS can be neglected compared with the first term since $T_+ \ll T_e$. For the collisional drag (third) term on the RHS, a constant mean-free path was employed, whereby the ion-neutral collision frequency as a function of ion velocity $\nu_+(u_+)$ was written as $\nu_+(u_+) = \nu_{+0}|u_+| / u_{+,th}$; $\nu_{i0}$ is a reference collision frequency at which the ion drift velocity equals the ion thermal speed $u_{+,th} = \sqrt{eT_+ / m_+}$. The fourth term on the RHS represents a "drag" in the sense that ions produced by ionization have negligible velocity, and have to be brought to the respective drift velocity . The ionization rate $R_i$ in Eq. 30 was calculated using the EEDF $f_0$ as

$$R_i(x) = N_0\sqrt{\frac{2e}{m}} \int_{\varphi(x)}^{\infty} \sigma_i(u)(\varepsilon - \varphi(x)) f_0(\varepsilon) d\varepsilon, \quad (23)$$

where $N_0$ is the gas density (assumed constant), $\sigma_i(u)$ is the ionization cross-section as a function of electron kinetic energy $u = \varepsilon - \varphi(x)$, not to be confused with velocity! An expression similar to Eq. (23) was used for computing the metastable ionization rate.

The boundary condition for ion velocity was set at the wall as

$$u_+ = -|u_b| = \left|\sqrt{eT_{eff}(0) / m_+}\right|, x = 0 \quad (24)$$

where the Bohm velocity $u_b$ was evaluated using the screening temperature at the edge. Due to symmetry, there was no flux at the discharge center (*x=L/2*).

For pressures below ~ 10 mTorr, a spatially-averaged (0-D) model can be employed for metastables,

$$\frac{dn_m}{dt} = -\frac{1}{L}\left(\frac{\gamma}{2-\gamma} n_m u_{m,th}\right) + \frac{2}{L}\int_0^{L/2}\left(\sum_j R_{mj}\right) dx \quad (25)$$

where $n_m$ is the metastable density and $u_{m,th}$ is the metastable thermal velocity. The summation on the right hand side is over all reactions accounting for production and loss of metastables in the gas phase. The first term on the right hand side represents the loss of metastables on the left boundary, wherein a flux expression [27] with deactivation probability $\gamma = 1$ was used.

## 3.0 METHOD OF SOLUTION

A modular approach was adopted in order to solve for the coupled EEDF, RF electric properties, and heavy species density and velocity profiles, in a self-consistent manner. The simulation consisted of three modules: an Electron Energy Distribution Function (EEDF) Module, a non-local electron kinetics (NLEK) module and a heavy species transport (HST) module. Figure 3 shows the interaction of the modules and the type of information exchanged among them. The NLEK module computed the non-local conductivity kernel *G(x,x')* (Eq. 12) and solved for the RF electric field (Eq. 10) and current density (Eq. 9) profiles given the peak electron density and the EEDF $f_0(\varepsilon)$. The RF electric field at the wall $E_0$ was adjusted to match the target total power. The RF field profiles were then used in the EEDF module to compute the energy diffusion coefficient $D_\varepsilon(\varepsilon)$ (Eq. 4) and solve for the distribution function $f_0(\varepsilon)$ (Eq. 3). The EEDF module provided the ionization (Eq. 23) and excitation rates, effective temperature ($T_{eff}$) profile (Eq. 20), and potential profile (Eq. 21). The effective temperature was used in the ion momentum equation (Eq. 22) and the boundary condition for ion velocity at the wall (Eq. 24). The ionization and excitation rates were used as source terms in the continuity equations for ions and metastables (Eqs. 15 and 25, respectively). The HST module provided, among other quantities, the ion (electron) density and potential $\varphi$ as a function of position. These were fed back to the NLEK module to calculate a new RF field profile. The calculation was repeated until convergence to a self-consistent solution.

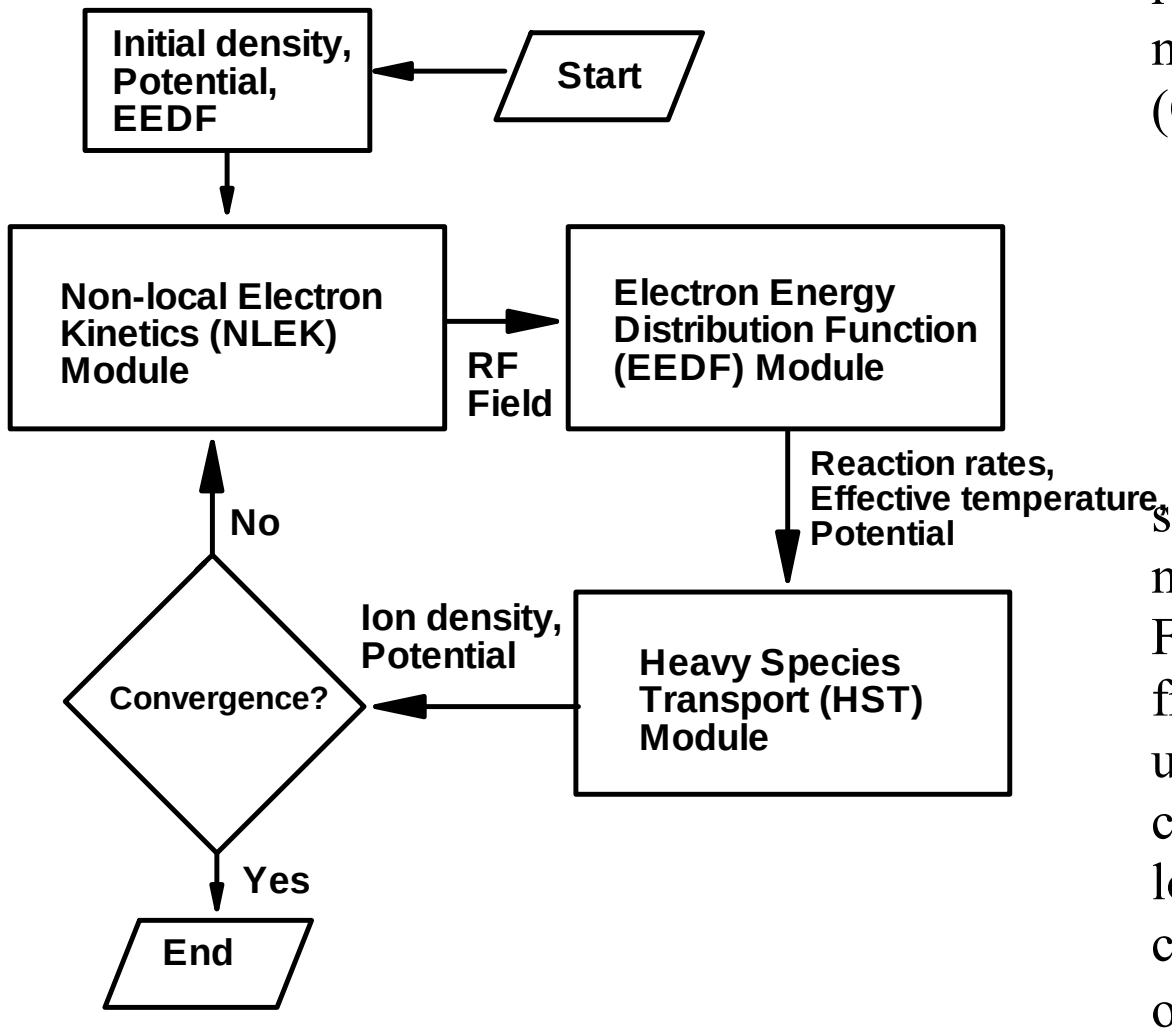

**Figure 3**: Flow diagram used for numerical simulation. Information exchanged between the modules is shown. Simulation cycled between the three modules until convergence.

The initial electron density profile was assumed to be a sine function peaking at the center of the reactor. The corresponding potential was computed assuming a uniform Maxwellian temperature of 2.5 V using Eq. (21). Convergence was declared when the potential profile changed by less than 0.1% (in the $L_2$ norm), which typically took about 30 iterations among the modules. The $L_2$ norm of the difference between the ion (computed by Eq. 15) and electron (computed by the EEDF, Eq. 3) density profiles was less than 0.1% at convergence.

In the NLEK module, Eq. (10) was discretized using second order central differencing for the left hand side. The integral on the right hand side was found using the trapezoidal rule after computing the conductivity kernel *G(x,x')* at each grid point. The grid spacing $\Delta x$ was chosen such that $\Delta x \ll l$, where, the "effective" mean-free path $l = V_T / \sqrt{\omega^2 + \nu^2}$. The discretized system of dense linear equations was then solved using a direct solver.

The equations in the HST module (Eqs. 15, 22 and 25) were solved simultaneously using a "staggered mesh" approach. The domain $0 \le x \le L/2$ was divided into the requisite number of cells. $Ar^+$ density was computed at cell centers, while $Ar^+$ velocity and flux were computed at the cell edges. A first order upwind scheme was used to compute the mass and momentum flux of ions at cell edges, naturally preserving the property of information propagating only outwards. The set of ordinary differential/algebraic equations resulting after discretization of Eqs. (15), (22) and (25) was integrated in time using Backward Difference Formulae methods [28] until a steady state was reached. Integration was performed to a real time of ~ 10 ms to ensure that metastables with the longest response time attain steady state. Calculations on a 933 MHz Intel Pentium 3 Windows NT Workstation took about 10 hrs for a converged run using the non-local conductivity (Eq. 9), and about 2 mins for a converged run using the local conductivity (Ohm's law, Eq. 7).

## 4.0 RESULTS AND DISCUSSION

Base-case parameter values used in the simulation are shown in Table II. Results in Figs. 4-9 are for a pressure of 1 mTorr and discharge frequency of 13.56 MHz. Results in Figs. 10-15 are for a pressure of 10 mTorr and discharge frequency of 13.56 MHz. In each case, profiles calculated using the non-local electron kinetics module (solid lines) are compared with profiles (dashed lines) obtained using the local approximation (Ohm's law) under the same discharge conditions and for the same (integrated) *total* power. Values of power correspond to a plate cross sectional area of 64π $cm^2$.

**Table II**: Parameter values used in the simulation

| Parameter | Value |
|---|---|
| Discharge length, $L$ | 5 cm |
| Ion temperature, $T_i$ | 0.026 V |
| Discharge cross sectional area, $A$ | 64π $cm^2$ |
| Reference ion collision frequency, $\nu_{i0}$ (at 1 mTorr, 300 K) | 15 kHz |
| Electron momentum transfer collision frequency, $\nu_{en}$ (@ 3.2 $10^{14}$ $cm^{-3}$) | 3 $10^7$ $s^{-1}$ |
| Gas Temperature | 300 K |
| Discharge frequency | 13.56 MHz |

### 4.1 Pressure =1 mTorr

Fig. 4 shows the EEDF as a function of *total* energy for the non-local (solid lines) and local (dashed lines) cases. For a pressure of 1 mTorr, the electron-neutral collision frequency is $\nu \sim 3\text{x}10^6$ s$^{-1}$, $\nu/\omega \sim 0.05$, and $\nu / \Omega_b \sim 0.1$. As a result, the energy diffusion coefficient (Eq. 4) exhibits a "knee" at ~ 1 V, indicating that the "temperature" of electrons with energies less than 1 V is significantly lower than that of electrons with energies greater than 1 V, i.e., low energy electrons are not heated as effectively. Such a knee is not displayed by the local limit of the energy diffusion coefficient. Consequently, the EEDFs for the local and non-local cases exhibit different behavior, especially for low energies and also for energies around the ionization threshold (15.76 V). Fig. 4 also shows a higher fraction of electrons just beyond the ionization threshold for the non-local case. The non-local and local models predict an effective temperature of ~ 6.5 V and ~ 5.5 V, respectively, for electrons near the discharge edge. This implies a difference in the effective electron mean free path near the edge (where the RF field is highest), which in turn leads to considerably different field and current density profiles, as discussed below.

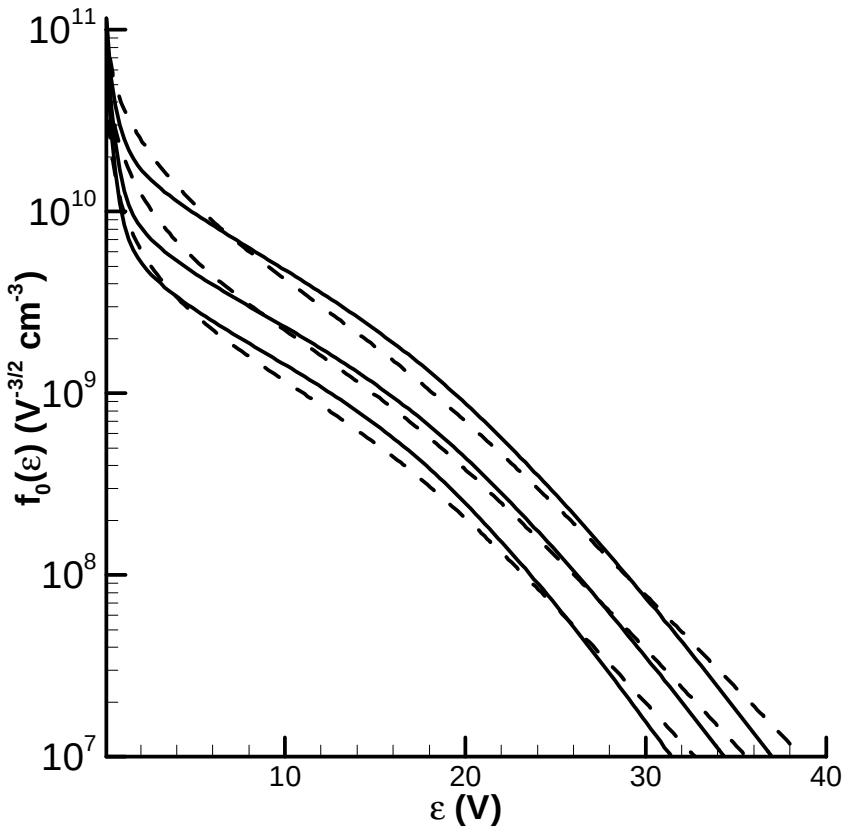

**Figure 4**: EEDF as a function of total energy for a pressure of 1 mTorr, discharge frequency of 13.56 MHz, and for three different powers. Results using the non-local electron kinetics model (solid lines) are compared with those using a local (Ohm's Law) model (dashed lines), under otherwise identical conditions. non-local (solid lines) and local (dashed lines) cases.

Fig. 5 shows the amplitude of the RF field as a function of position for total powers of 50, 100 and 200 W. The field was normalized by its value at the edge (*x=0*). Non-local behavior (non-monotonic field) is evident, especially at higher powers. The ratio of the effective electron mean free path $l$ to the skin depth in the cold plasma approximation $\delta_0$ is given by (see Eq. 1)

$$\frac{l}{\delta_0} = \frac{\omega_{pb} V_T}{\omega c} \frac{1}{\left(1 + \frac{\nu^2}{\omega^2}\right)^{3/4}} \tag{26}$$

where the electron plasma frequency $\omega_{pb}$ is based on the electron density at the sheath edge. For a non-Maxwellian EEDF, the electron thermal speed was defined as $V_T = \sqrt{2eT_{eff}(0)/m}$, where $T_{eff}(0)$ is the effective electron temperature at the sheath edge.

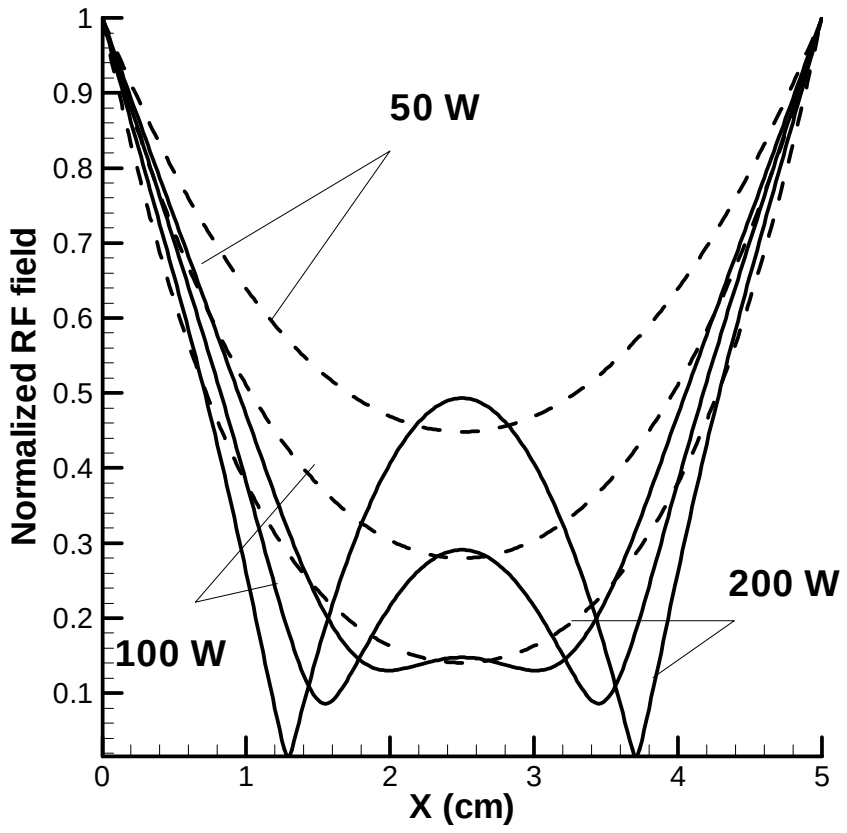

**Figure 5**: Normalized amplitude of the RF field for the same conditions as in Figure 4. Results using the non-local electron kinetics model (solid lines) are compared with those using a local (Ohm's Law) model (dashed lines), under otherwise identical conditions.

At 1 mTorr, the electron-neutral collision frequency is much smaller than the RF frequency, $\nu << \omega$, and the effective electron mean free path $l \simeq V_T / \omega \sim 1.65$ cm. At low power (50 W) $\delta_0 \sim 3$ cm, which is larger than $l$, and behavior is rather local. At high power (200 W), however, the skin depth reduces to 1.5 cm suggesting that a large fraction of electrons can contribute to the RF current outside the skin layer (non-local behavior). This is clearly seen in Fig. 6, which shows a much larger current density at the discharge center for the non-local case (the normalization factor for current density $J_0$ was defined as $J_0 = E_0 / \omega \mu_0 L^2$). Also, the peak in current density is much more pronounced, and much lower currents exist near the boundary in the non-local case. In this case, contributions to the current near the boundary can only come from electrons with energy larger than the sheath edge potential, $\varphi_{sh}$, i.e., a relatively small number of electrons.

The corresponding power deposition profiles are shown in Fig. 7. The non-local model predicts drastically different power deposition profiles compared to the local case. The maximum in the non-local power deposition profile is shifted away from the wall and, for higher powers, shows a pronounced feature near the center of the reactor. The local power deposition profile peaks near the wall and then decays monotonically towards the discharge center. This behavior can be explained noting that, in the local case, the phase difference between the current and the RF field is constant ($\Delta\phi = \tan^{-1}(\nu/\omega)$). Thus, the power deposition profile depends only on the magnitudes of the local conductivity and RF field (Ohm's law). In the non-local case however, $\Delta\phi$ is not constant. In fact, high-energy electrons coming from the discharge edge are out of phase with the RF field outside the skin layer (close to the center), and lose energy to the field, leading to negative power deposition. Negative power deposition has been observed experimentally in low pressure inductively coupled discharges [29].

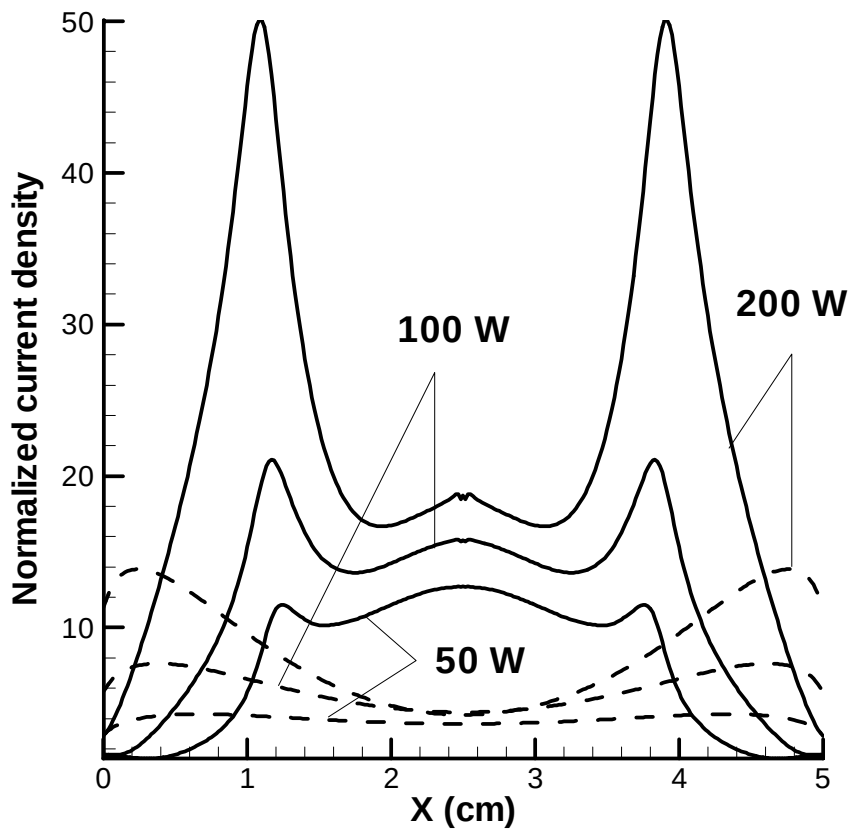

**Figure 6**: Normalized current density profiles for the same conditions as in Figure 4. Results using the non-local electron kinetics model (solid lines) are compared with those using a local (Ohm's Law) model (dashed lines), under otherwise identical conditions.

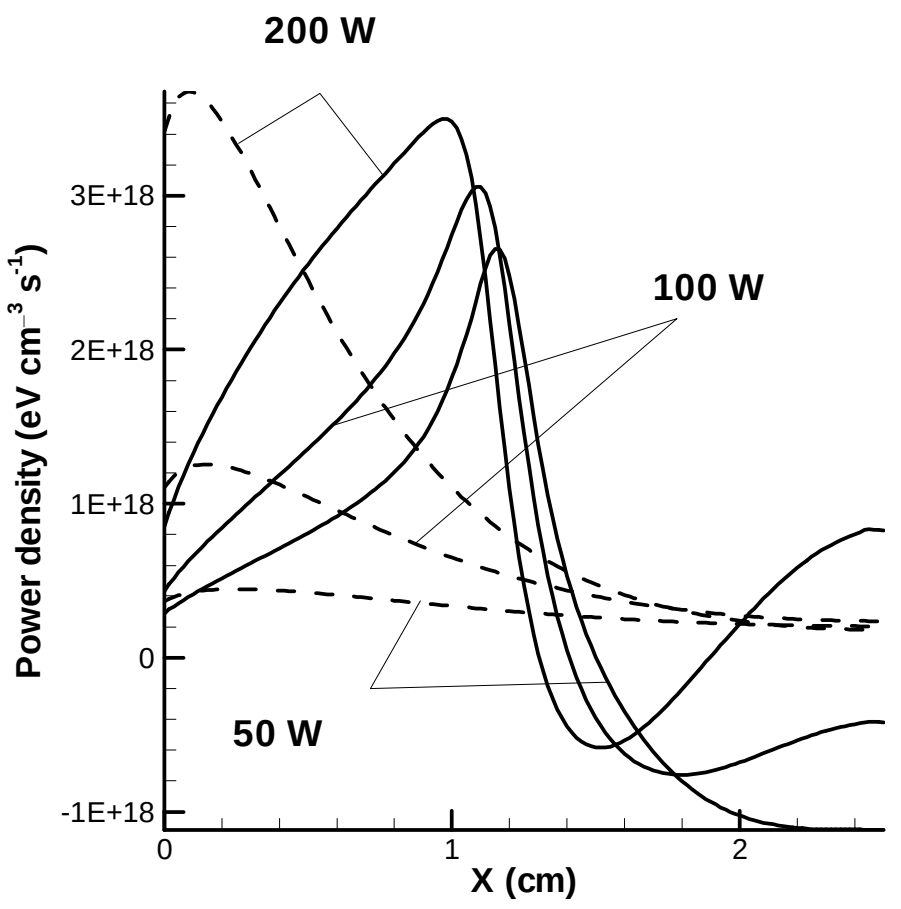

**Figure 7**: Power density profiles for the same conditions as in Figure 4. Results using the non-local electron kinetics model (solid lines) are compared with those using a local (Ohm's Law) model (dashed lines), under otherwise identical conditions.

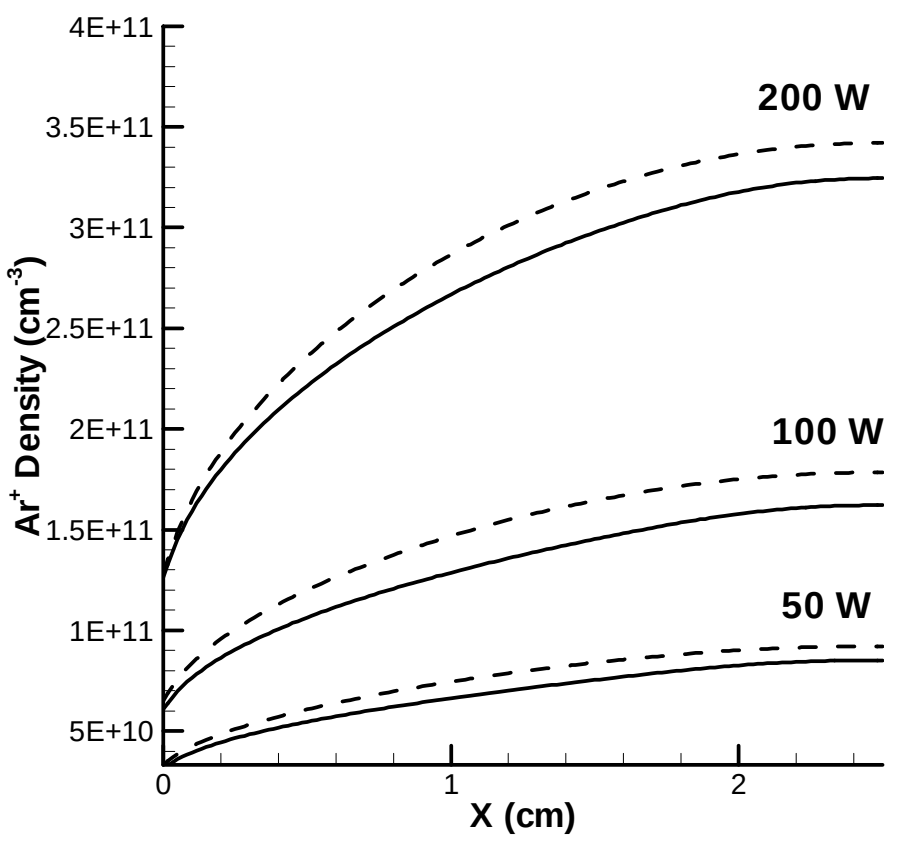

**Figure 8**: Positive ion density profiles for the same conditions as in Figure 4. Results using the non-local electron kinetics model (solid lines) are compared with those using a local (Ohm's Law) model (dashed lines), under otherwise identical conditions.

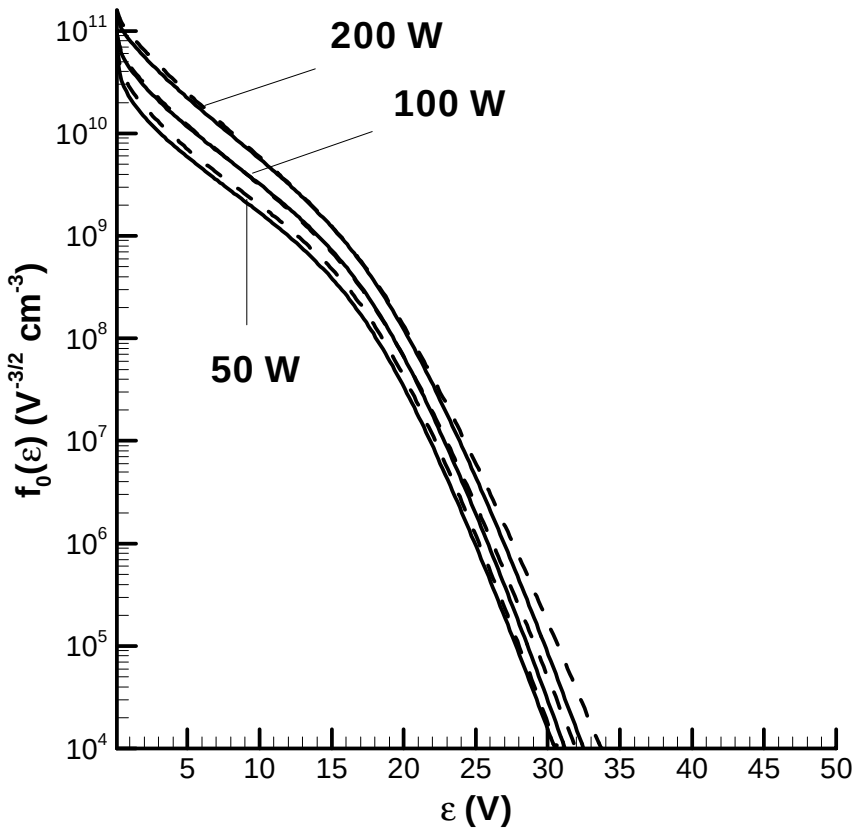

**Figure 9**: EEDF as a function of total energy for a pressure of 10 mTorr, discharge frequency of 13.56 MHz, and for three different powers. Results using the non-local electron kinetics model (solid lines) are compared with those using a local (Ohm's Law) model (dashed lines), under otherwise identical conditions.

The corresponding positive ion density profiles are shown in Fig. 8. The positive ion density is determined by production through ionization (e.g., Eq. 23), and loss at the boundary (Eq. 24). The latter depends on the effective electron temperature at the boundary, $T_{eff}$. The values of $T_{eff}$ at the boundary for the non-local and local cases were significantly different (6.5 and 5.5 eV, respectively) as the EEDFs differ considerably for energies below the ionization threshold (see Fig. 4). Furthermore, the rate of ionization (which depends mainly on the tail of the distribution function ($\varepsilon > 15.76$ V) since ground state ionization dominates) was found to be marginally higher for the non-local EEDF. These two trends counterbalance each other resulting in comparable ion densities for the non-local and local cases. The maximum deviation of ~ 5% occurs at a power of 200 W.

### 4.2 Pressure=10 mTorr

Fig. 9 shows the EEDF as a function of total energy for the non-local (solid lines) and local (dashed lines) cases. For a pressure of 10 mTorr, the electron-neutral collision frequency is $3x10^7$ $s^{-1}$, comparable to the discharge frequency ($f$=13.56 MHz , $\omega$=8.52x $10^7$ $s^{-1}$). In addition, $\nu > \Omega_b$ ($\Omega_b$ is the electron bounce frequency). As a result, the energy diffusion coefficient (Eq. 4) tends to the local value. Consequently, the EEDFs for the local and non-local cases are very similar. Furthermore, since the electron density at these pressures is sufficiently high, electron-electron collisions tend to thermalize the distribution, leading to an EEDF close to Maxwellian. In fact, electrons at the bulk and at the edge have an effective temperature of ~3 V and ~3.5 V, respectively (weakly dependent on power). In contrast, for 1 mTorr, electrons at the bulk and edge have an effective temperature of ~6.5 eV and ~2 eV, respectively (for the non-local EEDF, see Fig. 4).

Fig. 10 shows the profiles of the amplitude of the RF field as a function of position for total power of 50, 100 and 200 W. The profiles are again normalized by the value of the field at the edge. A transition from local (monotonic decay from edge to center) to slightly non-local (non-monotonic decay) behavior is observed as power is increased. However, the effect of non-locality is not pronounced at this pressure as $l / \delta_0 < 1$ for all powers. For example, at low power (50 W) $\delta_0$ ~ 2.9 cm, larger than the effective electron mean free path $l$ ~ 1.2 cm. Hence most electrons are confined within the skin layer, and the local approximation is valid. As power increases, the skin depth decreases (becoming ~ 1.5 cm for a power of 200 W), while the effective mean-free path remains approximately constant. This causes some energetic electrons to escape from the skin layer during an RF cycle, resulting in slightly non-local behavior.

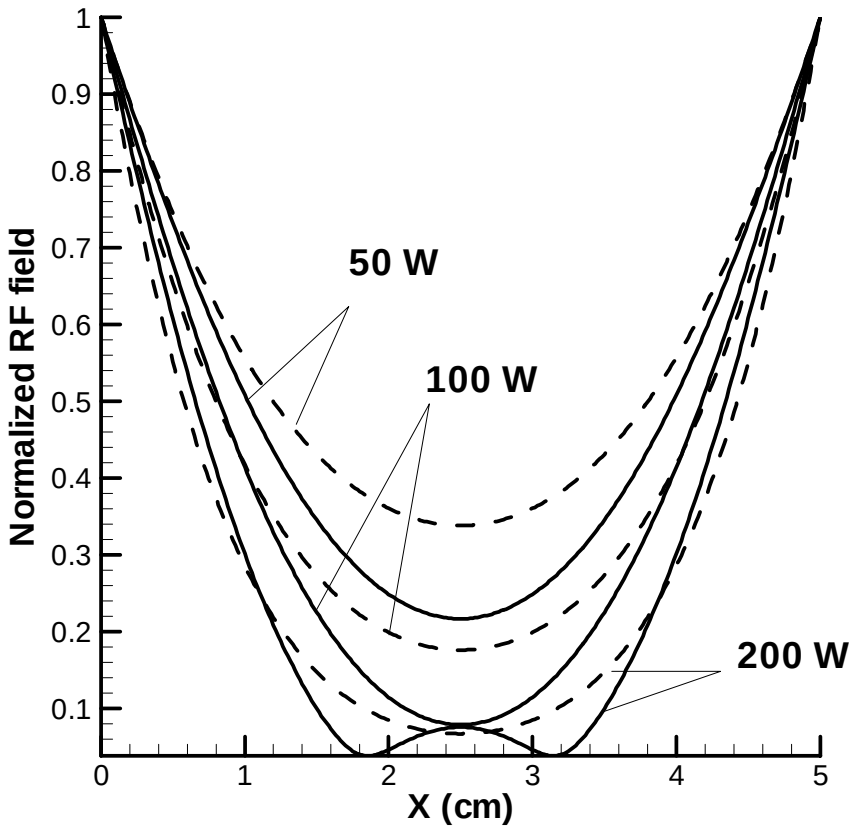

**Figure 10**: Normalized amplitude of the RF field for the same conditions as in Figure 9. Results using the non-local electron kinetics model (solid lines) are compared with those using a local (Ohm's Law) model (dashed lines), under otherwise identical conditions.

The normalized current density profile is shown in Fig. 11 for the same conditions as in Fig. 10. The normalization factor $J_0$ was defined as $J_0 = E_0 / \omega\mu_0 L^2$. Since non-locality is not very pronounced at this pressure, the current density profiles for the local and non-local cases are similar, as power is varied from 50 W to 200 W. As expected, the peak in the current density occurs further away from the boundary for the non-local case (but still within the skin layer). The low current density near the boundary is due to lower electron density in that region, while the low current density near the discharge center is due to a weaker RF electric field at the center.

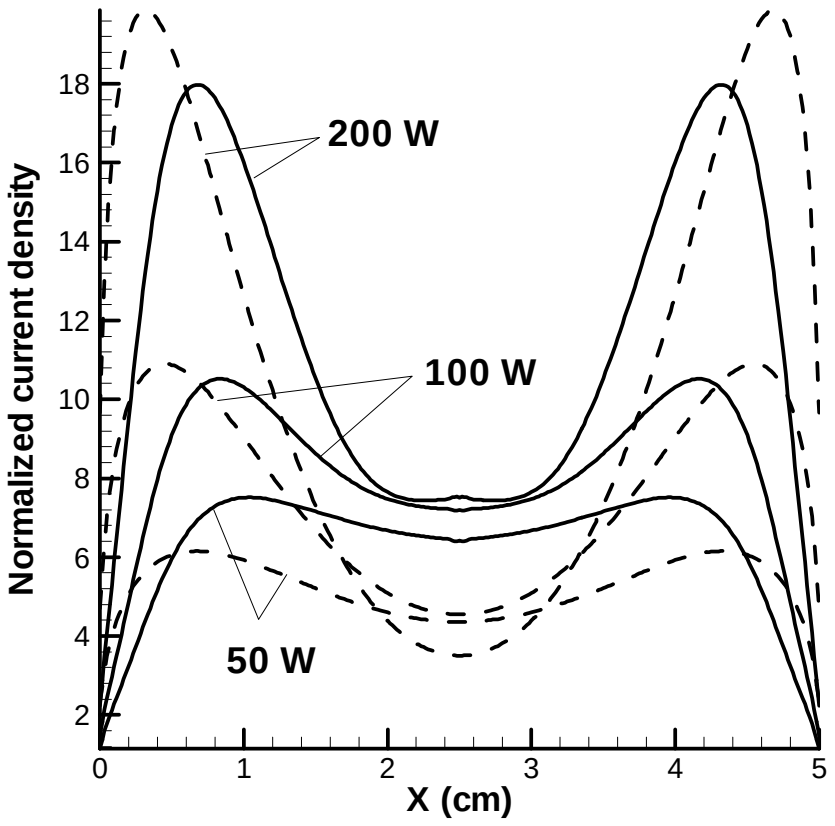

**Figure 11**: Normalized current density profiles for the same conditions as in Figure 9. Results using the non-local electron kinetics model (solid lines) are compared with those using a local (Ohm's Law) model (dashed lines), under otherwise identical conditions.

The corresponding power deposition profiles are shown in Fig. 12. Power deposition reaches a maximum within the skin layer and decays towards the center of the discharge. The profiles for the local and non-local cases are similar, except for the slight negative power deposition near the discharge center for 200 W (for which non-local behavior is relatively more pronounced). The peak of power deposition occurs closer to the boundary, well within the skin layer, when compared to the profiles at 1 mTorr (Fig. 7). This suggests that energy absorption is mainly collisional. If the pressure were to increase to 20 mTorr, the power deposition profiles for the local and non-local cased would be even closer to one another.

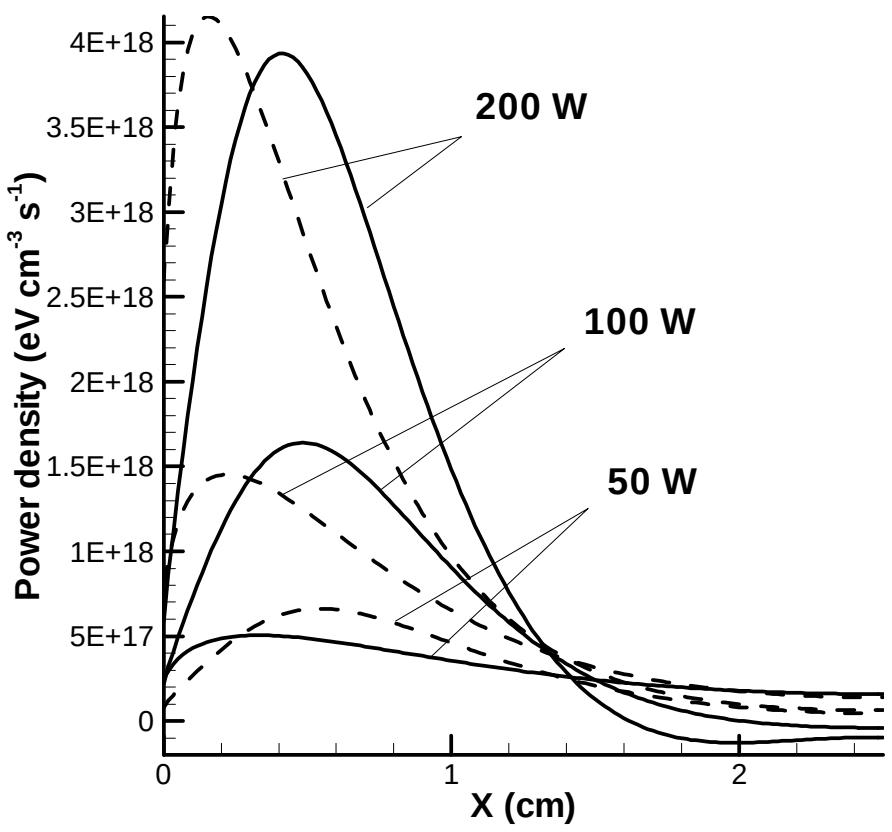

**Figure 12**: Power density profiles for the same conditions as in Figure 9. Results using the non-local electron kinetics model (solid lines) are compared with those using a local (Ohm's Law) model (dashed lines), under otherwise identical conditions.

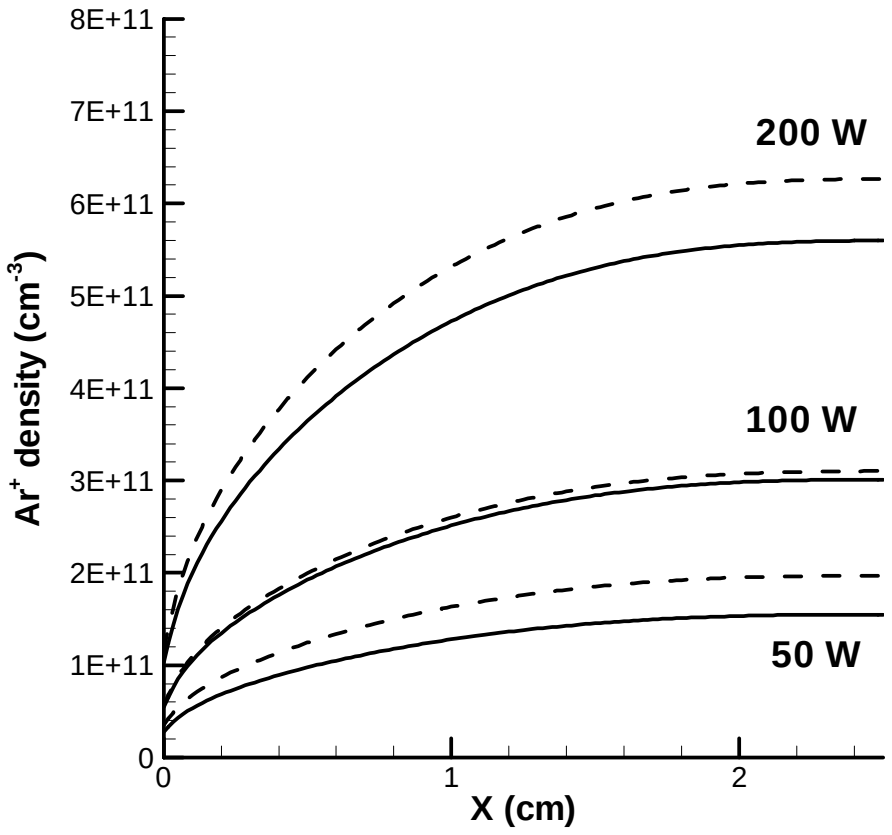

**Figure 13**: Positive ion density profiles for the same conditions as in Figure 9. Results using the non-local electron kinetics model (solid lines) are compared with those using a local (Ohm's Law) model (dashed lines), under otherwise identical conditions.

The corresponding positive ion density profiles are shown in Fig. 13. When comparing the non-local and local cases, the effective electron temperature at the boundary is very similar (3.5 V and 3.4 V, respectively) and the rate of ionization is also similar due to comparable "tails" of the EEDF (Fig. 9). As a result, the ion densities for the non-local and local cases are very similar.

Differences in RF field and current profiles in local and non-local cases may be characterized by computing the surface impedance [15] of the discharge, defined by $Z = -i\omega\mu_0 E_y(0) / \left[dE_y / dx\right]_{x=0}$. Table IV shows a comparison of surface impedance predicted by the non-local and local models. The local model grossly underestimates the surface impedance, especially for low pressure (1 mTorr) and high power (200 W) for which non-locality is more pronounced.

**Table IV**: Real part of surface impedance for local and non-local models.

| Pressure | Power | Surface Impedance ($\Omega$) (Non-local) | (Local) |
|---|---|---|---|
| 1 mTorr | 50 W | 0.163 | 0.045 |
| | 100 W | 0.354 | 0.029 |
| | 200 W | 0.633 | 0.020 |
| 10 mTorr | 50 W | 0.520 | 0.312 |
| | 100 W | 0.450 | 0.203 |
| | 200 W | 0.418 | 0.141 |

### 4.3 Comparison with experiments

Figures 14-17 show comparisons between experimental data [30] and simulation predictions using the non-local (Figs. 14 and 16) and local (Figs. 15 and 17) models. The Ar ICP chamber was 19.8 cm in (inside) diameter and 10.5 cm in length. The coil current driving frequency was 6.78 MHz. The model described in Section 2 was modified to handle this asymmetric discharge. The boundary condition for the normalized RF field was set to zero on the boundary opposite the coil (metal wall). Eq. (12) was also modified, assuming a Maxwellian EEDF, with an electron temperature obtained experimentally [30]. The ion (electron) density was set to the value measured experimentally [30]. The normalization factor $E_0$ for the RF field at the wall on the coil-side was obtained by matching the simulated integrated power deposition with the experimental value of power.

Figs. 14a and 14b show reasonable agreement between predicted (using the non-local model) and measured RF field and current density profiles. The field decreases to a minimum and then goes through a hump. The qualitative features are captured with the model, although the precise location of the minimum is not captured. This may be due to the fact that the actual discharge is 2-D, while the model is only 1-D. The predictions of the local model (Fig 15) are in serious discrepancy with the measurements. Table III shows skin depths predicted by the local and non-local models compared with the experimental values. For the non-local model, the predicted skin depth is closer to the experimental values for higher powers. For the local model, the predicted skin depth deviates more at higher powers, due to increased non-locality as power increases. Clearly, the non-local model is in better agreement with measurements.

Similar comparisons for 10 mTorr are shown in Figs. 16 and 17. The non-local model again does a better job in capturing the experimental profiles, although the local model is not as bad as it was for 1 mTorr. The local profiles in Figs. 17a and 17b show reasonable agreement with experiment for low powers when non-locality is less prominent.

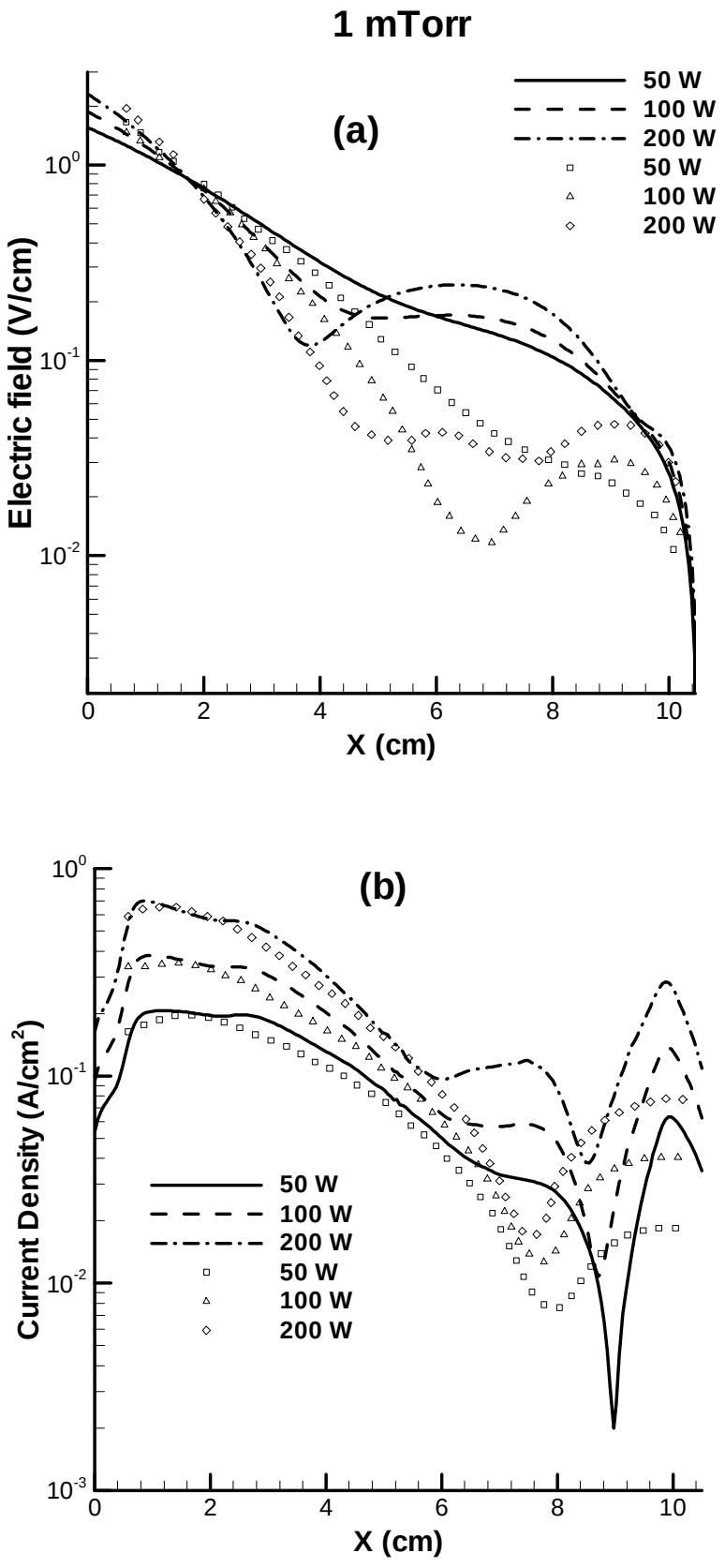

**Figure 14**: Comparison between experimental data [30] and simulation predictions using a non-local model of RF electric field (a), and current density (b) profiles for a pressure of 1 mTorr.

Table III: Comparison between experimental skin depth [30] and skin depth predicted by the local and non-local models.

| Pressure | Power | Skin Depth (cm) | | |
|---|---|---|---|---|
| | | Experiment | Simulation (Non-local) | Simulation (Local) |
| | 50 W | 1.96 | 2.14 | 1.72 |
| 10 mTorr | 100 W | 1.63 | 1.72 | 1.24 |
| | 200 W | 1.48 | 1.51 | 0.99 |
| | 50 W | 2.18 | 2.04 | 2.04 |
| 1 mTorr | 100 W | 2.04 | 1.98 | 1.44 |
| | 200 W | 1.83 | 1.85 | 1.02 |

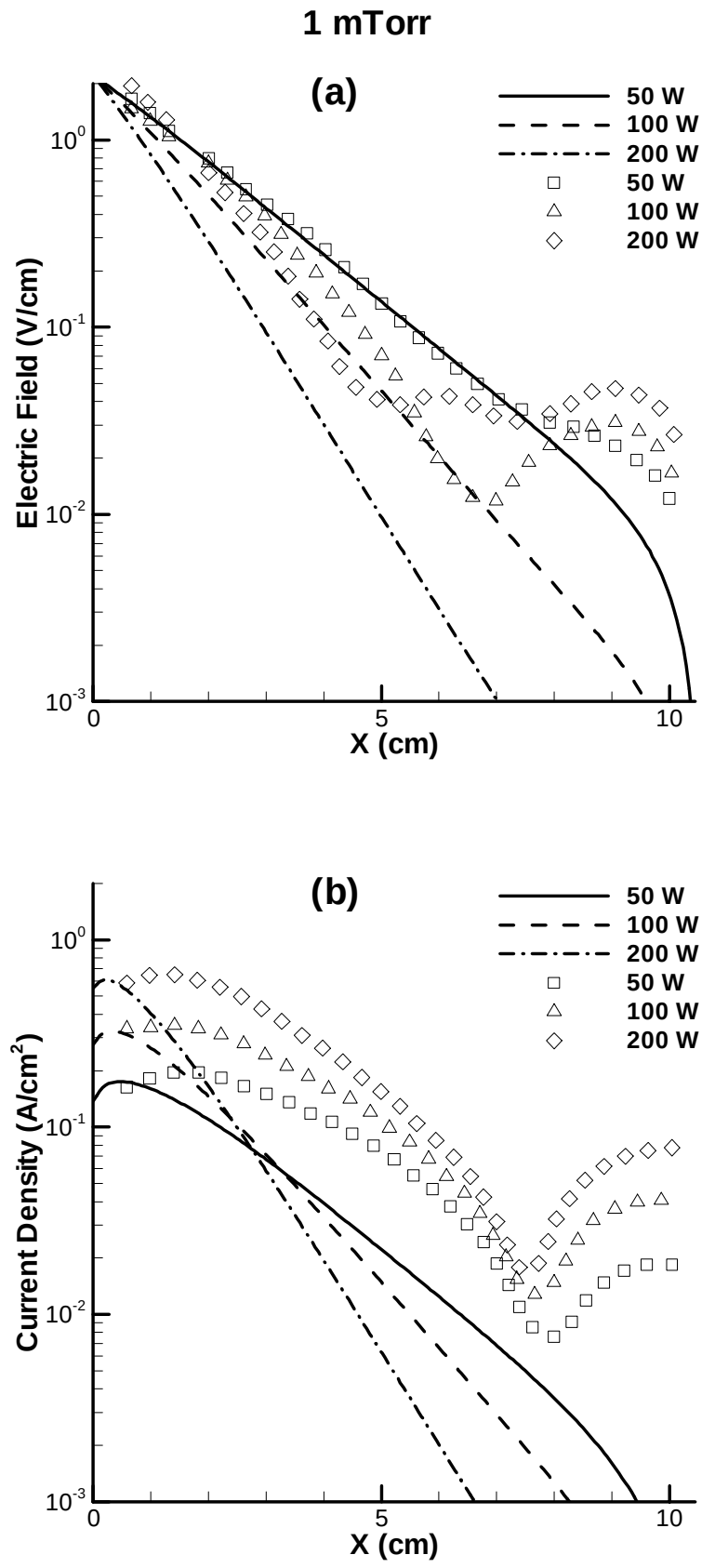

**Figure 15**: Comparison between experimental data [30] and simulation predictions using a local model of RF electric field (a), and current density (b) profiles for a pressure of 1 mTorr.

## 5.0 CONCLUSIONS

A self-consistent 1-D model was developed to study the effects of non-local electron conductivity on power absorption and plasma density profiles in a planar inductively coupled argon discharge at low pressures (≤ 10 mTorr). The model consisted of three modules: (1) an electron energy distribution function (EEDF) module to compute the non-Maxwellian EEDF, (2) a non-local electron kinetics module to predict the non-local electron conductivity, RF current, electric field and power deposition profiles in the non-uniform plasma, and (3) a heavy species transport module to solve for the ion density and velocity profiles as well as the metastable density. The self-consistent simulation predicted the RF electric field, power deposition, electron energy distribution function, and ion density profiles. Results using the non-local electron conductivity model were compared with predictions of a local theory (Ohm's law), under otherwise identical conditions.

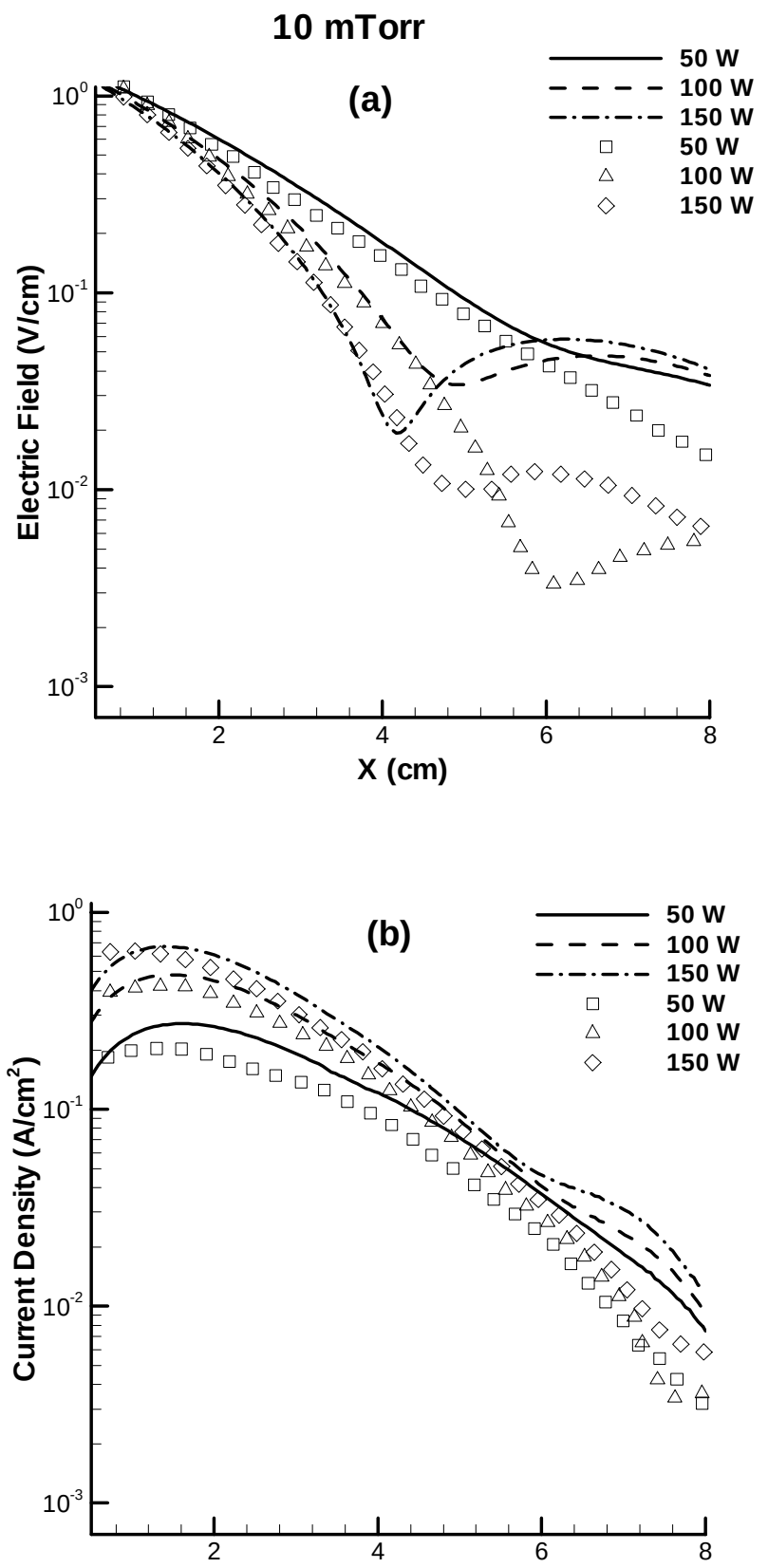

**Figure 16:** Comparison between experimental data [30] and simulation predictions using a non-local model of RF electric field (a), and current density (b) profiles for a pressure of 10 mTorr.

The non-local power deposition profile peaked further away from the wall, and exhibited a pronounced secondary peak near the center of the discharge, where negative power deposition was also observed, especially for higher total power (higher electron density and thinner skin layer). In contrast, the local power deposition profile peaked closer to the wall and then decayed monotonically towards the discharge center. This difference is due to a warm plasma effect: high energy electrons originating from inside the skin layer can contribute to current outside the skin layer, near the discharge center. Since these electrons can be out of phase with the RF field, negative power deposition can be observed near the discharge center, especially for higher powers (more non-locality). Non-local effects were particularly pronounced at 1 mTorr, for which the electron collision frequency was much smaller than both the discharge frequency and the bounce frequency.

The EEDFs predicted by the local and non-local models were found to be similar at 10 mTorr. At this pressure, the electron-neutral collision frequency is larger than the bounce frequency, and electron heating is substantially collisional. At a pressure of 1 mTorr, however, the EEDF predicted by the non-local model revealed that electrons with energies greater than 1 V were heated more efficiently, when compared to the local case. This resulted in substantial differences in the EEDFs predicted by the local and non-local models.

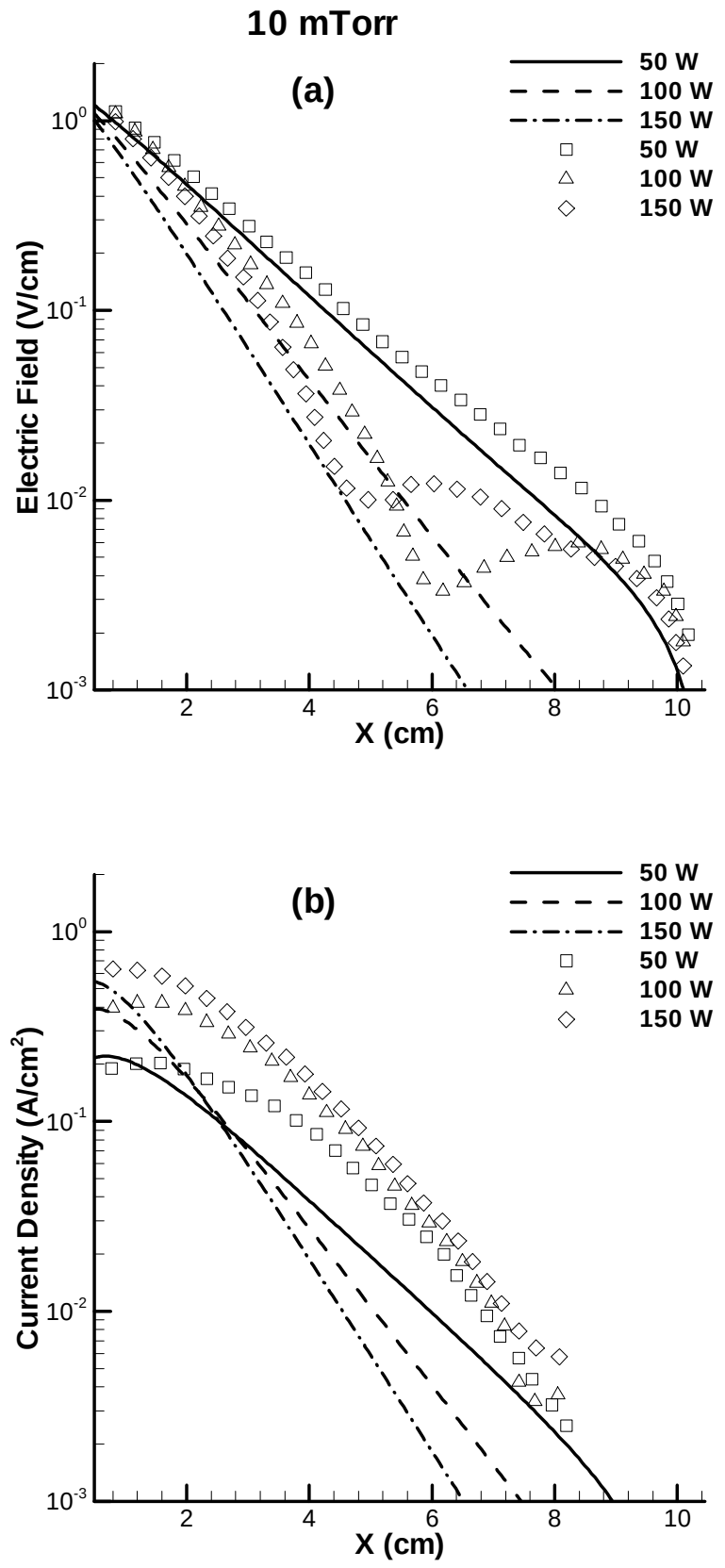

**Figure 17**: Comparison between experimental data [30] and simulation predictions using a local model of RF electric field (a), and current density (b) profiles for a pressure of 10 mTorr.

Although the RF electric field, current, and power deposition profiles predicted by the local and non-local models were drastically different (especially at 1 mTorr), the ion (and electron) density profiles were very similar, differing no more than 10%. The larger ionization rate predicted by the non-local EEDF, was counterbalanced by a larger effective temperature at the wall (more ion escape by the Bohm flux), resulting in essentially identical ion density profiles. This finding suggests that, for the prediction of plasma density profiles, a local electron conductivity model (Ohm's law), which is computationally much faster, may be used even at very low pressures.

The above comparison between non-local and local conductivity models was made for the actual EEDF as determined self-consistently. A separate paper [31] reports the effect of the EEDF on discharge electrical properties and plasma density profiles. It is shown that the plasma density can differ by as much as 70% when a Maxwellian EEDF is used instead of the actual EEDF at pressures below 10 mtorr.

## ACKNOWLEDGEMENTS

BR and DJE are thankful to the National Science Foundation (CTS 0072854) for financial support. IDK acknowledges support of the Princeton Plasma Physics Laboratory University Research Support Program.

## REFERENCES

1 J. Hopwood, C. R. Guarnier, S. J. Whitehair, and J. J. Cuomo, *J. Vac. Sci. and Technol. A* **11**, 152 (1993).

2 J. H. Keller, J. C. Foster, and M. S. Barnes, *J. Vac. Sci. and Technol. A* **11**, 147 (1993).

3 P. L. G. Ventzek, T. J. Sommerer, R. J. Hoekstra and M. J. Kushner, *J. Vac. Sci. and Technol. A* **12**, 461 (1994); T. Panagopoulos, V. Midha, D. Kim and D. J. Economou, *J. Appl. Phys*., **91**, 2687 (2002).

4 R. B. Piejak, V. A. Godyak, and B. M. Alexandrovich, *Plasma Sources Sci. and Technol*. **3**, 169 (1994).

5 D. J. Economou, *Thin Solid Films*, **365**, 348 (2000); A. P. Paranjpe, *J. Vac. Sci. and Technol. A* 12, 2487 (1994).

6 M. M. Turner, *Phys. Rev. Lett.* **71**, 1844 (1993).

7 V. A. Godyak, R. B. Piejak, and B. M. Alexandrovich, *Phys. Rev. Lett*. **80**, 3264 (1998).

8 G. Cunge, B. Crowley, D. Vender, and M. M. Turner, *J. of Appl. Phys*. **89**, 3580 (2001).

9 E. S. Weibel, *Phys. Fluids* **10**, 741 (1967).

10 M. A. Liberman, B. E. Meierovich, and L. P. Pitaevskii, *Sov. Phys. JETP* **35**, 904 (1972).

11 S. M. Dikman and B. E. Meierovich, *Sov. Phys. JETP* **37**, 835 (1973).

12 R. A. Demirkhanov, I. Kadysh, and Yu. S. Khodyrev, *Sov. Phys. JETP* **19**, 791 (1964).

13 H. A. Blevin, J. A. Reynolds, and P. C. Thonemann, *Phys. of Fluids* **13**, 1259 (1970).

14 H. A. Blevin, J. A. Reynolds, and P. C. Thonemann, *Phys. of Fluids* **16**, 82 (1973).

15 K. C. Shaing, and A. Y. Aydemir, *Phys. of Plasmas* 4, 3163 (1997).

16 N. S. Yoon, S. S. Kim, C. S. Chang, and Duk-In Choi, *Phys. Rev. E* **54**, 757 (1996).

17 I. D. Kaganovich, V. I. Kolobov and L. D. Tsendin, *Appl. Phys. Lett*. **69**, 3818 (1996).

18 Chin Wook Chung, K.-I. You, S. H. Seo, S. S. Kim, and H. Y. Chang Physics of Plasmas, **8**, 2992 (2001).

19 I. D. Kaganovich, *Phys. Rev. Lett*. **82**, 327 (1999).

20 V. I. Kolobov, and D. J. Economou, *Plasma Sources Sci. Technol*. **6**, 1 (1997).

21 V. A. Godyak, V. I. Kolobov, *Phys. Rev. Lett*. **81**, 369 (1998).

22 Yu. M. Aliev, I. D. Kaganovich and H. Schluter, *Phys. Plasmas* **4**, 2413 (1997).

23 I. D. Kaganovich, submitted to Physics of plasmas (2001).

24 U. Kortshagen, C. Busch and L. D. Tsendin, *Plasma Sources Sci. Technol*. **5**, 1 (1996).

25 D. P. Lymberopoulos and D. J. Economou, *J. Appl. Phys.* **73**, 3668 (1993).

31 B. Ramamurthi, D. J. Economou, and I. D. Kaganovich, submitted to Plasma Sci. Technol. (2002).

26 K. U. Riemann, *J. Phys. D* **24**, 493 (1991); S. V. Berezhnoi, I. D. Kaganovich and L. D. Tsendin, *Plasma Sources Sci. Technol*. **7**, 268 (1998); S. V. Berezhnoi, I. D. Kaganovich and L. D. Tsendin, *Plasma Physics Reports* **24**, 556 (1998).

27 P. J. Chantry, *J. Appl. Phys.* **62**, 1141 (1987); N.N. Semenoff, Acta Phys. USSR, **18**, 93 (1943).

28 G. D. Byrne and A. C. Hindmarsh, *J. Comp. Phys.* **70**, 1 (1987).

29 V. A. Godyak and V. I. Kolobov, *Phys. Rev. Lett*. **79**, 4589 (1997).

30 V. A. Godyak and R. B. Piejak, *J. Appl. Phys*. **82**, 5944 (1997).